%% file: outflow_main.tex
\documentclass[fleqn,usenatbib]{mnras}
\input{additional_tex/pack}        
\input{additional_tex/definitions} 
\input{additional_tex/additional_def} 

\graphicspath{{plots/}} 

\defcitealias{chevalier_clegg:1985}{CC85}
\def\citeCC85{\citetalias{chevalier_clegg:1985}}
\defcitealias{Fujimoto19}{F19}
\def\citeF19{\citetalias{Fujimoto19}}

\title[~\CII halos in high-z galaxies]{Outflows and extended \CII halos in high redshift galaxies}

\author[Pizzati et al.]{E. Pizzati$^{1}$\thanks{\href{mailto:elia.pizzati@sns.it}{elia.pizzati@sns.it}},
A. Ferrara$^{1,2}$,
A. Pallottini$^{1,2}$,
S. Gallerani$^{1}$,
L. Vallini$^{3}$,
D. Decataldo$^{1}$,
\newauthor S. Fujimoto$^{4}$
\\
$^{1}$ Scuola Normale Superiore, Piazza dei Cavalieri 7, 56126 Pisa, Italy\\
$^{2}$ Centro Fermi, Museo Storico della Fisica e Centro Studi e Ricerche \quotes{Enrico Fermi}, Piazza del Viminale 1, Roma, 00184, Italy\\
$^{3}$ Leiden Observatory, Leiden University, PO Box 9500, 2300 RA Leiden, The Netherlands\\
$^{4}$ The Cosmic Dawn Center, Niels Bohr Institute, University of Copenhagen, VIbenshuset 4. sal, Lyngbyvej 2, 2100 Copenhagen, Denmark}

\date{Accepted XXX. Received YYY; in original form ZZZ}

\pubyear{2020}

\begin{document}
\label{firstpage}
\pagerange{\pageref{firstpage}--\pageref{lastpage}}
\maketitle

\begin{abstract}
Recent stacked ALMA observations have revealed that normal, star-forming galaxies at $z\approx 6$ are surrounded by extended ($\approx 10$ kpc) \CII emitting halos which are not predicted by the most advanced, zoom-in simulations. We present a model in which these halos are the result of supernova-driven cooling outflows. Our model contains two free parameters, the outflow mass loading factor, $\eta$, and the parent galaxy dark matter halo circular velocity, $v_c$. The outflow model successfully matches the observed \CII surface brightness profile if {$\eta = 3.20 \pm 0.10$ and $v_c = 170 \pm 10 \kms$}, corresponding to a dynamical mass of $\approx 10^{11}\,\msun$. The predicted outflow rate and velocity range are $128 \pm 5\,\msun {\rm yr}^{-1}$ and $300-500 \kms$, respectively. We conclude that: (a) extended halos can be produced by cooling outflows; (b) the large $\eta$ value is marginally consistent with starburst-driven outflows, but it might indicate additional energy input from AGN; (c) the presence of \CII halos requires an ionizing photon escape fraction from galaxies $\fesc \ll 1$. The model can be readily applied also to individual high-$z$ galaxies, as those observed, e.g., by the ALMA ALPINE survey now becoming available.  
\end{abstract}




\section{Introduction}

The advent of radio-interferometers such as ALMA and NOEMA has offered for the first time the opportunity to investigate the internal structure of galaxies located deep into the Epoch of Reionization (EoR, redshift $z>6$).
These studies are now nicely complementing large scale near-infrared surveys which have successfully characterised the evolution of the rest-frame galaxy UV luminosity functions, star formation, stellar-build up history, and size evolution, thus building a solid statistical characterisation of these earliest systems up to $z\approx 10$. We defer the interested reader to the recent review by \citet{Dayal18} and references therein.   

Thanks to Far Infrared (FIR) emission lines such as \CII 158$\mu$m, \OIII 88$\mu$m, CO from various rotational levels, and dust continuum we are rapidly improving our understanding of the small-scale, internal properties and assembly history of galaxies in the EoR, including their interstellar medium and relation to star formation \citep{capak2015,carniani2017}, gas dynamics \citep{agertz:2015apj, pallottini2017,Hopkins18}, spatial offsets \citep{inoue2016,Laporte17,carniani2017,carniani:2018}, dust and metal enrichment \citep{capak2015,Tamura:2019,behrens2018, Knudsen:2017, Laporte17}, the molecular content \citep{vallini:2018,dodorico:2018}, interstellar radiation field \citep{stark2015,pallottini:2019}, and outflows \citep{gallerani:2018}.

Due to its brightness (it is one of the major coolant of the ISM) the \CII $^2P_{3/2} \rightarrow ^2P_{1/2}$ fine-structure transition at $1900.5469\,\mathrm{GHz}$ ($157.74 \,\mu\mathrm{m}$) has been routinely used as a work-horse for the investigations. A sample of tens of $z>6$ galaxies is now available, providing solid starting point for morphological and dynamical studies of these systems.

One of these studies \citep[][\citeF19 hereafter]{Fujimoto19} has combined 18 galaxies $5.1 < z < 7.1$ by applying the stacking technique in the uv-visibility plane to ALMA Band 6/7 data. Quite surprisingly, this study found (at 9.2$\sigma$-level) that the radial profiles of the \CII surface brightness is significantly ($\approx 5\times$) more
extended than the HST stellar continuum and ALMA dust continuum. In absolute terms the detected halo extends out to approximately $10 \,\mathrm{kpc}$ from the stacked galaxy center. This discovery parallels the extended emission found in a more massive, $z\sim 6$ quasar host, galaxy \citep{cicone2015}, where the [CII] emission is detected up to 20-30 kpc, while the FIR emission does not exceed 15 kpc. Similar results have been also found in stacked \citet{ginolfi:2019} and individual \citep{fujimoto:2020} galaxies using data of the ALPINE survey (ALMA LP, PI: O. Lefevre, \citealt{lefevre:2019}, \citealt{Faisst:2020}, \citealt{bethermin:2019}).
Moreover, since the galaxies considered by \citeF19 have SFR between 10 and 100 $\msun$, their discovery suggests that a cold carbon gas halo universally exists even around early \quotes{normal} galaxies. \citet{Rybak20} found a significantly extended \CII emission around SDP.81, a $z=3.042$ gravitationally lensed dusty star-forming galaxy. They report that $\approx 50$ per cent of \CII emission arises outside the FIR-bright region of the galaxy.

The previous findings resonate with similar existing evidences of extended Ly$\alpha$ halos around high-$z$ galaxies. By using 26 spectroscopically confirmed Ly$\alpha$-emitting galaxies at $3<z<6$, \citet{Wisotzki16} found that most of these low-mass systems show the presence of extended Ly$\alpha$ emission that are 5-15 times larger than the central UV continuum sources as seen by HST. In a follow-up work, \citet{Wisotzki18} demonstrated that the projected sky coverage of Ly$\alpha$ halos of galaxies at $3<z<6$ approaches 100\%. Ly$\alpha$ intensity mapping experiments confirm this scenario. \citet{Kakuma19} identify very diffuse Ly$\alpha$ emission with 3$\sigma$ significance at > 150 comoving kpc away from Lyman Alpha Emitters at $z=5.7$, i.e. beyond the virial radius of star-forming galaxies whose halo mass is $10^{11}\,\msun$. These independent evidences for extended halos pose their existence on very solid grounds.

The discovery of extended \CII halos around early galaxies raise three challenging physical questions: (a) by what means has carbon (and presumably other heavy elements) been transported to these large distances from the galactic centre where it was produced by stellar nucleosynthesis; (b) how can carbon atoms remain in a singly ionized state in the presence of the cosmic UV background produced by galaxies and quasars, rather than being found in higher ionization states as routinely observed in low-density, unshielded environments such as e.g. the Ly$\alpha$ forest \citep{Dodorico13}; (c) what is the carbon mass required to explain the observed \CII emission \citep{vallini2015,pallottini2015,kohandel:2019} at these high redshifts? 
Such questions make clear that the origin, structure and survival of \CII halos represent a formidable problem in galaxy evolution.

The existence of extended \CIIion halos might also affect profoundly our views on metal enrichment of the intergalactic medium \citep{Dodorico13,Meyer19,Becker19}, and have an impact on future intensity mapping \citep{Yue15, Yue19} experiments (for an overview, see \citealt{Kovetz17} and references therein) specifically targeting \CII signal from the galaxy population predominantly responsible for cosmic reionization. 
Extended halos, in fact, might leave a very specific signature in the 1-halo term of the \CII power spectrum clustering signal. 

The problem is particularly severe as even the most physically-rich, zoom-in simulations \citep{pallottini2017b, Arata18} fail to reproduce the observed \CII surface brightness. These independent studies almost perfectly agree in predicting a \CII halo profile that drops very rapidly beyond $2\,\mathrm{kpc}$, and, at $8\,\mathrm{kpc}$ from the centre, has a \CII luminosity $\approx 10\times$ below that observed.
Different scenarios have been proposed to explain the presence of abundant \CII emission at large galactocentric distances: satellite galaxies, outflows, cold accreting streams. The last scenario (cold accreting streams that flow from the CGM into the galaxy) finds little support from theoretical considerations and observations, since there is no compelling evidence for their presence, and because they are expected to be metal poor. The emission from a population of faint galaxy satellites is in principle a good candidate for solving the problem.
However, while faint satellite galaxies are indeed seen in simulations, they do not provide a sufficient luminosity to account for the emission \citep{pallottini:2019}. Moreover, this answer appears to be in contrast with observations, since as shown in \citeF19 (Sec. 4.3), the ratio between \CII emission and the total SFR surface density is not compatible with the hypothesis of dwarf galaxies.

Thus, it appears that the outflow scenario is the most promising explanation. According to this hypothesis, the halos represent an incarnation of outflows driven by powerful episodes of star formation and/or AGN activity occurring in high-$z$ galaxies. In this context, we note the $z\approx 6$ quasar host galaxy showing the extended halo detected by \citet{cicone2015}, is also known to have a powerful AGN-driven outflow. Evidences for the presence of outflows around normal galaxies at $z\approx 6$ are further suggested by ALMA observations in a sub-set of the F19 sample \citep{gallerani:2018}, and now further supported by the ALPINE Large Program \citep{ginolfi:2019}.
Fast outflows have been tentatively identified in $z=5-6$ galaxies also using deep Keck metal absorption line spectra \citep{Sugahara19}. According to both observations and detailed simulations, outflows often present a multi-phase structure composed by different outflow modes \citep{murray2011, hopkins2014, muratov2015, heckman2017}. Hot modes ($T\approx 10^{6-7}\,\mathrm{K}$) are often fast and highly-ionized, while cold modes ($T\approx10^{2-4}\,\mathrm{K}$) are neutral and slower. Cold modes are often formed by radiative cooling of the hot gas outflowing from the galaxy. Different works highlight the role of this catastrophic cooling in regulating feedback mechanisms in super-star clusters \citep{ Silich:2004,gray2019catastrophic}, and galaxies \citep{Wang:1995, Thompson15, Thompson16, McCourt:2015, sarkar:2015, Scannapieco:2017, Schneider:2018, Gronke&Oh:2020}, and suggest that the outflow mass budget is likely to be dominated by cold gas.
An outflow that undergoes catastrophic cooling could transport carbon in singly ionized form away from the galaxy, and the \CII emission could arise from suitable conditions of high density and low temperature. 

Here we explore this idea using a semi-analytical model for a cooling outflow and simulating the resulting \CII emission in order to compare it directly with observations from \citeF19. We conclude that outflows represent a possible answer to the origin of the observed \CII halos, and we show that -- in spite of the simplifications required to implement this idea -- the results are robust and provide at least a reliable framework for a more detailed work.

The paper is organized as follows: in Sec. \ref{sec_adiabatic_outflows} and Sec. \ref{sec_cooling_outflow} we present the hydrodynamical model for the outflow making different physical assumptions; in Sec. \ref{sec_structure_of_the_outflow} we discuss the resulting structure for the outflows in terms of the loading parameters; in Sec. \ref{sec_CII_emission} we work on modelling the \CII emission; in Sec. \ref{sec_comparison_with_data} we compare the results from our model with the observational data and with other previous works; conclusions are given in Sec. \ref{sec_conclusions}.

\section{Adiabatic outflows} \label{sec_adiabatic_outflows}

To model gas outflows from galaxies, we start by considering the classical study by \citet[][hereafter {\citeCC85}]{chevalier_clegg:1985}. Among the many necessary simplifying assumptions made by the authors, the most critical one for our study is that the flow is adiabatic and cools only by expansion. Therefore, in the next Sec. we will increment the {\citeCC85} model by including both gravity and radiative cooling terms, following also similar work by \citet{Thompson16}. 

Our aim is to derive physically-motivated density, velocity and temperature radial profiles of the outflow as a function of model parameters. These quantities will form the basis for the prediction of \CII luminosity that we present in Sec. \ref{sec_CII_emission}.

The {\citeCC85} model describes a spherically symmetric, hot, and steady wind that drives energy and mass -- injected by stellar winds and supernovae (SNe) -- out of the galaxy. Energy and mass are uniformly deposited by the central stellar cluster in a region of radius $R$ at a constant rate, equal to $\dot E$ and $\dot M$, respectively.

We relate these quantities to the star formation rate (SFR) via two efficiency parameters, $\alpha$ and $\eta$, such that
\begin{subequations}\label{eq:def_Mdot_Edot}
\begin{align}
\dot{M}&=\eta\, \mathrm{SFR}\\
\dot{E}&=\alpha \nu E_0\, \mathrm{SFR}, 
\end{align}
\end{subequations}
where $E_0=10^{51}\, \rm erg$ is the SN explosion energy, and $\nu = 0.01\, \msun^{-1}$ is the number of SNe per unit stellar mass formed. The mass loading factor, $\eta$, heavily affects the gas density, and thus the general behaviour of the system. The dependence of the physical variables on $\alpha$ is not as strong, and to a first approximation it can be fixed. For this reason, we have decided to set $\alpha=1$ \citep[chosen accordingly to outflow observations by][]{strickland2009supernova}, and retain $\eta$ as the only parameter in our model. 

Outside the injection region ($r>R$), mass, momentum, and energy are conserved; the wind expands against the vacuum (we neglect the presence of the interstellar medium). Additional simplifications include neglecting the presence of viscosity and thermal conduction. The latter is generally a fair assumption, apart from some extreme regimes involving low values of $\eta$ \citep[see][in particular Sec. 2.2 therein]{Thompson16}.

As already mentioned, in this Sec. we neglect both radiative cooling and gravity. The first assumption is equivalent to the condition that the cooling time, $\tau$, largely exceeds the advection time (i.e. a gas parcel is removed from the system before it is able to radiate). Neglecting gravity implies that the outflow velocity is much larger than centrifugal velocity, $v_{c}$, from the system. We will release these assumptions in the next Sections. 

\begin{figure}
	\centering
	\includegraphics[width=0.49\textwidth]{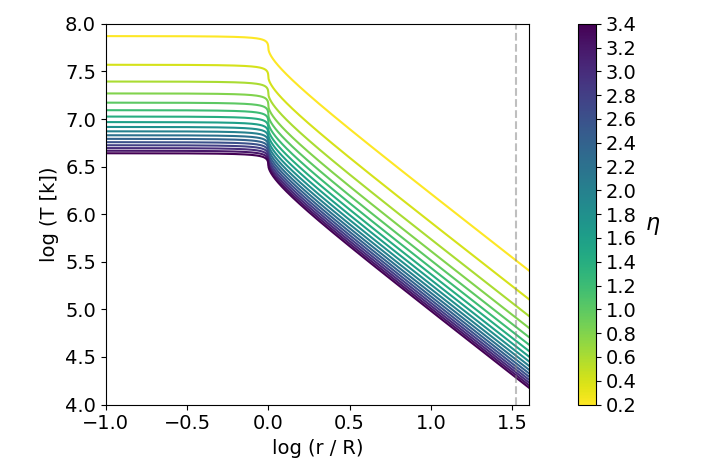}
	\caption{Outflow radial temperature (T) as a function of the radius (r) in the adiabatic model. The curves are calculated for $R=300$ pc and SFR$=50 \, \msun \rm yr^{-1}$. Different colors indicate different values of the mass loading factor ($\eta$). The gray dashed line indicates the distance $r=10\,\mathrm{kpc}$.
	\label{fig:temp_CC85}
	}
\end{figure}
With these hypothesis, we write the relevant hydrodynamical equations assuming a spherically symmetric, steady-state flow as follows:
\begin{subequations}
\begin{align}
&\frac{1}{r^2}\der{}{r}(r^2v\rho)=q  \label{eq:eul1}\\
&\rho v\der{v}{r}= -\der{p}{r} -vq \label{eq:eul2}\\
&\frac{1}{r^2}\der{}{r}\bigg[r^2\rho v\bigg(\rho \frac{v^2}{2}+\frac{\gamma}{\gamma -1}\frac{p}{\rho}\bigg)\bigg]=Q, \label{eq:eul3}
\end{align}
\end{subequations}
where $\rho, v, p$ are the gas density, velocity and pressure; the mass input rate $q$ and energy input rate $Q$, assumed to be constant, take the form
\begin{equation}
\begin{cases} q=\frac{3\dot{M}}{4\pi R^3}\,, \,\,\,r\le R\\ q=0\,, \,\,\,r> R \end{cases}
\qquad \begin{cases} Q=\frac{3\dot{E}}{4\pi R^3}\,, \,\,\,r\le R\\ Q=0\,, \,\,\,r> R \end{cases}
\end{equation}
These equations are complemented by an adiabatic equation of state (EoS) with index $\gamma = 5/3$.

Solutions can be found by imposing the appropriate boundary conditions: $v(0)=0$, $p(r\rightarrow+\infty)=\rho(r\rightarrow+\infty)=0$, and matching the derivatives of the solutions at $r=R$ (critical point). Using the Mach number -- $\mathcal{M}=v/c_s$ where $c_s^2=\gamma p/\rho$ is the gas sound speed -- the conditions can be expressed as
\begin{align}
&\bigg(\frac{3 \gamma + 1/\mathcal{M}^2}{1+3 \gamma}\bigg)^{-(3\gamma+1)/(5 \gamma+1)}\bigg(\frac{\gamma-1+2/\mathcal{M}^2}{1+\gamma}\bigg)^{(\gamma+1)/(2(5\gamma+1))} = \frac{r}{R} \label{eq:solinn}\\
&\mathcal{M}^{2/(\gamma-1)}\bigg(\frac{\gamma-1+2/\mathcal{M}^2}{1+\gamma}\bigg)^{(\gamma+1)/(2(\gamma-1))} = \bigg(\frac{r}{R}\bigg)^2 \label{eq:solout}\,,
\end{align}
where eq. \ref{eq:solinn} (eq. \ref{eq:solout}) applies to the inner, $r<R$ (outer, $r>R$) region. 

From the Mach number and the boundary conditions, we can directly obtain the profiles for $v$, $n$, $P$, and $T$. In Fig. \ref{fig:temp_CC85} we show the outflow temperature profile for different values of the mass loading parameter $\eta$ in the range 0.2-3.4; note that we use the values $R=300\,\mathrm{pc}$ and SFR$=50 \,\msun \rm yr^{-1}$. For $r<R$ the temperature is roughly constant at $10^{7-8}$ K, with the exact value depending on $\eta$: more mass-loaded outflows are cooler. Beyond $R$, the temperature drops purely due to adiabatic cooling following the characteristic behavior $T\propto r^{-4/3}$. 

We clearly see that an adiabatic outflow cannot account for the observed \CII halo emission. In fact, $T>10^5$ K within the central 10 kpc for all models. At these temperatures \CIIion ions are still largely collisionally ionized to higher ionization states, with the consequent suppression of the 158 $\mu$m line emission. It is then necessary to introduce cooling effects (and gravity) in the model. This is discussed in the next Sec..

\section{Cooling outflows} \label{sec_cooling_outflow}

We follow \citet[][]{Thompson16}, and rewrite the hydrodynamical equations introducing the net (i.e. cooling $-$ heating) cooling function, $\Lambda(T,n,r)$, and an external gravitational potential. We assume that the gravitational potential, $\Phi$, is provided by the dark matter halo, whose density distribution is approximated by an isothermal sphere for which $\rho(r) \propto r^{-2}$.  In principle, one should also include the gravitational contribution due to the baryonic component in the galaxy disk. However, the disk potential decreases as $r^{-1}$ while the DM halo potential increases logarithmically with distance from the center of the galaxy. This implies that beyond a kpc scale (where the physics of the outflow becomes more interesting) the disk contribution is completely irrelevant, and therefore we neglect it.

The gravitational potential is parameterized via the galaxy circular velocity 
\begin{align}
v_c=\sqrt{\frac{G M(r)}{r}}.
\label{eq:v_esc}
\end{align}
Since for an isothermal sphere $M(r)\propto r$, then $v_c = {\rm const}$. We use $v_c=175 \kms$ as the fiducial value for the galaxies in the \citeF19 sample, but we also explore the dependence of the results on this parameter in Sec. \ref{sec_v_esc0}. The boundary conditions at $r=R$ are obtained by integrating the \citeCC85 equations in the inner region. 

Within the inner region we adopt the standard \citeCC85 model which neglects radiative losses. This is justified by the fact that the temperature (Fig. \ref{fig:temp_CC85}) is approximately constant around $10^{7-8}\,\mathrm{K}$: at these temperatures the cooling time is far greater than the advection time. In addition, we neglect gravity effects in the inner region as they affect only very marginally the boundary conditions \citep{bustard2016}.
Writing explicitly the solutions for the physical variables in the inner region, we cast the boundary conditions in the form:
\begin{subequations}\label{eq:bc}
\begin{align}
&\rho(R)=\frac{\sqrt{2}}{4\pi}\,\frac{\dot{M}^{3/2}}{\dot{E}^{1/2}}\frac{1}{R^2}\propto {\rm SFR}\, \eta^{3/2},\\
&p(R)=\frac{3\sqrt{2}}{40\pi}\,\frac{\dot{M}^{1/2}\dot{E}^{1/2}}{R^2}\propto {\rm SFR} \,\eta^{1/2},\\
&v(R)=\frac{1}{\sqrt{2}}\,\frac{\dot{E}^{1/2}}{\dot{M}^{1/2}}\propto \eta^{-1/2}\,,\label{eq:v_boundary}
\end{align}
\end{subequations}
where the r.h.s terms are obtained using eqs. \ref{eq:def_Mdot_Edot}.

We now focus on the outer region where $q=Q=0$. There, the mass, momentum and energy conservation equations read 
\begin{subequations}
\begin{align}
&\frac{1}{r^2}\der{}{r}(r^2v\rho) =0, \label{eq:neweul1}\\
&\rho v\der{v}{r} = -\der{p}{r} - \rho \der{\Phi}{r},\\
&\bigg[\frac{1}{T}\der{T}{r} - (\gamma-1)\frac{1}{\rho}\der{\rho}{r}\bigg] v k_B T = - (\gamma-1) n \Lambda. \label{eq:neweul2}
\end{align}
\end{subequations}
Combining the three equations we get a first order system of ODE that can be integrated numerically to solve for the variables $\rho$, $v$, and $T$. These equations can be written in terms of the flow Mach number $\mathcal{M}$, the gravitational Mach number $\mathcal{M}_g=v_c/c_s$, and the cooling time $\tau = k_B T/n\Lambda$ as:
\begin{subequations}
\begin{align}
&\der{\log\rho}{\log r}=2 \left(\frac{\mathcal{M}^2-\mathcal{M}_g^2/2}{1-\mathcal{M}^2}\right)+ \frac{r}{\lambda_c}\left(\frac{1}{1-\mathcal{M}^2}\right) \label{eq:cooling_model_equations_n}\\
&\der{\log v}{\log r}=\left(\frac{\mathcal{M}_g^2-2}{1-\mathcal{M}^2}\right)+\frac{r}{\lambda_c}\left(\frac{1}{1-\mathcal{M}^2}\right) \label{eq:cooling_model_equations_v}\\
&\der{\log T}{\log r}={2(\gamma-1)}\left(\frac{\mathcal{M}^2-\mathcal{M}_g^2/2}{1-\mathcal{M}^2}\right)-\frac{r}{\lambda_c}\left(\frac{1-\gamma\mathcal{M}^2}{1-\mathcal{M}^2}\right)\,, \label{eq:cooling_model_equations_T}
\end{align}
\label{eq:cooling_model_equations}
\end{subequations}
where $\lambda_c=[\gamma/(\gamma-1)] v \tau$ is the cooling length.

\subsection{Radiation fields}\label{Radfields}

In order to solve eqs. \ref{eq:cooling_model_equations} it is necessary to specify the functional form of the net cooling function $\Lambda(T,n,r)$. This function is affected by the presence of a UV radiation field in two ways: (a) heating due to photoelectric effect on gas and/or dust; (b) photoionization of cooling species which result in a lower emissivity of the gas. Both effects tend to decrease the value of $\Lambda$ at a given temperature; therefore they should be carefully modelled in order to reliably predict the emission properties of the outflow.

There are two main sources of UV radiation in the galactic halo environment: (a) stars in the parent galaxy, and (b) the cosmic UV background (UVB) produced by galaxies and quasars on cosmological scales.  While the stellar flux decreases with distance $r$ from the galaxy, the UVB can be considered to a good approximation as spatially constant at a given redshift. The relative intensity of the two radiation fields depends also on the fraction of ionizing photons produced by stars that are able to escape into the halo, i.e. the so-called escape fraction, $\fesc$.

If $\fesc$ is large, we show below that the galactic radiation field dominates the UVB up to distances that are considerably larger than those ($\approx 10$ kpc) relevant here. However, local and high-$z$ observations \citep[for a review see][]{Dayal18, inoue2006escape} indicate that most systems are characterized by very low ($\simlt$ few percent) escape fractions. Given the present uncertainties we consider the case $\fesc=0$ as the fiducial one, but we also explore the implications of $\fesc=0.2$, the value usually invoked by most reionization studies \citep{Mitra15,Robertson15,Mitra18} to bracket all possible configurations. We note that if the properties of \CII halos turn out to be very sensitive to $\fesc$, they might be used as a novel way to measure $\fesc$ at early times.

To precisely evaluate the heating and ionization effects produced by the presence of radiation fields it is necessary to compute the corresponding H and He photoionization rates, as well as the photodissociation of \HH molecules, a key cooling species, by Lyman-Werner (LW, 912-1108 \AA) photons. We concentrate on this task in the next two Sec.s.

\subsubsection{Galactic flux}

We use the data tables from \code{Starburst99} \citep{leitherer1999} to get the specific luminosity, $L_\nu$ of the galaxy (stars + nebular emission). We choose a \citet{salpeter1955} IMF between $1$ and $100\,\msun$, using Geneva tracks \citep{schaerer1993}. Luminosities are computed for a continuous star formation rate of SFR $=50 \, \msun \rm yr^{-1}$ (fiducial value). The specific ionizing photon rate from the galaxy at radius $r$ and frequency $\nu$ is then:
\begin{align}
\mathcal{\dot N}_\nu=\frac{L_\nu}{h \nu} \fesc \,.
\end{align}
The corresponding photoionization rate for the \textit{i}-species (\textit{i} =H, He, C) is 
\begin{equation}
\Gamma_i=\int_{\nu_{T,i}}^{+\infty}\frac{\mathcal{\dot N}_\nu}{4\pi r^2}\alpha^i_\nu\mathrm{d}\nu \,,
\label{eq:photo_integration}
\end{equation}
where $\alpha^i_\nu$ is the photoionization cross-section of a given element, and the integration is performed from the ionization threshold at frequency $\nu_{T,i}$. We use the following fit for $\alpha^i_\nu$: 
\begin{equation}
\alpha_\nu=\alpha_\mathrm{T}\bigg[b\bigg(\frac{\nu}{\nu_\mathrm{T}}\bigg)^{-a}+(1-b)\bigg(\frac{\nu}{\nu_{\mathrm{T}}}\bigg)^{-a-1}\bigg]\,\, \mathrm{for}\,\, \nu > \nu_{\mathrm{T}}\,. \label{eq:crosssec}
\end{equation}
The adopted values of ($\alpha_T, \nu_T, a, b$) for the three species are given in Tab. \ref{tab:params}. For H and He we obtain
\begin{subequations}
\begin{align}
&\Gamma_{\mathrm{H}}(r) = 2.73\times 10^{-7} \, \bigg(\frac{\mathrm{kpc}}{r}\bigg)^2 \fesc \,\, \mathrm{s}^{-1},\\
&\Gamma_{\mathrm{He}}(r)=8.85\times 10^{-8}\, \bigg(\frac{\mathrm{kpc}}{r}\bigg)^2 \fesc\,\, \mathrm{s}^{-1}\,.
\end{align}
\label{eq:photo_profiles}
\end{subequations}

\begin{table}
\centering
\begin{tabular}{ c c c c c }
\hline
 Species & $\nu_\mathrm{T}$ $(10^5 \, \mathrm{cm}^{-1})$ & $\alpha_\mathrm{T}$ $(10^{-18} \,\mathrm{cm}^{2})$ & $a$ & $b$\\
\hline
\hline
 H & 1.097 & 6.3 & 2.99 & 1.34\\
 He & 1.983 & 7.83 & 2.05 & 1.66\\
 \CIion & 0.909 & 12.2 & 2.0 &3.35\\
 \CIIion & 1.97 & 4.60 & 3.0 & 1.95\\
 \hline
\end{tabular}
\caption{Photoionization cross-section parameters for H, He, and C entering eq. \ref{eq:crosssec}. Data from \citet{tielens2005book}.
\label{tab:params}
}
\end{table}

Finally, we compute the \HH photodissociation rate by LW photons. To this aim we use the relation given by \citet{anninos1997} linking the radiation field specific intensity at the LW band center (12.87 eV) with the photo-dissociation rate
\begin{align}
\Gamma_{\rm{H2}} = 1.38 \times 10^9 \, \mathrm{s}^{-1} \, \bigg(\frac{J_\nu(h\Bar{\nu}=12.87 \, \mathrm{eV})}{\mathrm{erg}\,\mathrm{s}^{-1}\,\mathrm{cm}^{-2}\,\mathrm{Hz}^{-1}\,\mathrm{sr}^{-1}}\bigg)\,; \label{eq:lw}
\end{align}
for our choice of the stellar population, and hence $J_\nu(h\Bar{\nu}=12.87 \, \mathrm{eV})$, this translates into:
\begin{align}
\Gamma_{\rm{H2}}(r)=1.42\times 10^{-8}\, \bigg(\frac{ \mathrm{kpc}}{r}\bigg)^2 \fesc \,\, \mathrm{s}^{-1}\,.
\label{eq:Lw_profile}
\end{align}
For simplicity we are assuming the same value of $\fesc$ for ionizing and non-ionizing (LW) photons. As the two escape fractions are influenced by different physical processes, they might however be slightly different.

\begin{figure*}
    \centering
    \includegraphics[width=1.025\textwidth,height=8.5cm]{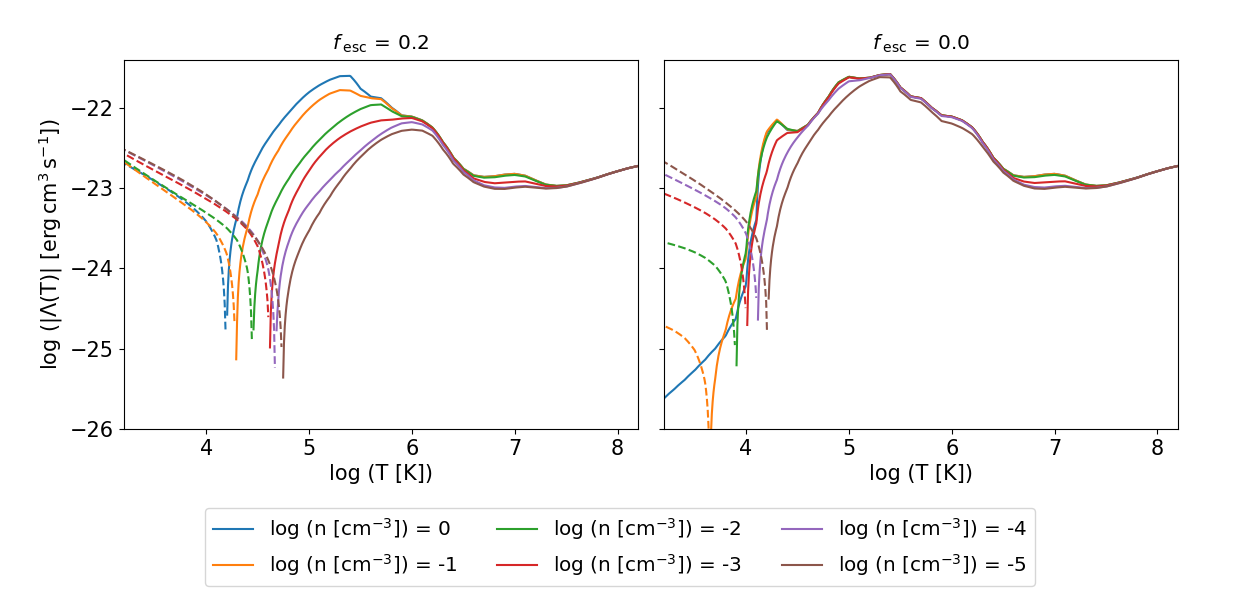}
    \caption{Net cooling function ($\Lambda(n,T,r)$) as a function of the temperature ($T$), for different values of the gas density ($n$).
    Note that the absolute value of $\Lambda$ is plotted: solid (dashed) lines represent positive (negative) values, i.e. net cooling (heating).
    The data for the cooling rates are taken from \citet{gnedin2012cooling}, and we have used as input the values of the photoionization and photodissociation rates $\Gamma_{\mathrm{H}}$, $\Gamma_{\mathrm{He}}$, and $\Gamma_{\mathrm{LW}}$ derived in Sec. \ref{Radfields}.
    {\it Left panel}: case for $\fesc=0.2$, in which the ionizing radiation field is given by the sum of the flux from the galaxy and the cosmic UVB at $z=6$. Results are shown at a galactocentric radius $r=1\,\mathrm{kpc}$. {\it Right}: case for $\fesc=0$. Ionizing radiation is only provided by the UVB.
    \label{fig:global_cooling}
    }
\end{figure*}

\subsubsection{Cosmic UV background}\label{par_UV_back}

We repeat the above calculation for the UVB at $z=6$ assuming a \citet[][]{haardt2012radiative} spectral shape and specific intensity, $J_\nu$, or
\begin{equation}\label{eq:UV_photo}
\Gamma_{\rm{UVB,i}}=4\pi \int_{\nu_{T,i}}^{+\infty}\frac{J_\nu}{h\nu} \alpha_\nu^i \mathrm{d}\nu \,.
\end{equation}
The integral gives the H and He photoionization rates, ($\Gamma_{\mathrm{UVB,H}}, \Gamma_{\mathrm{UVB,He}})= (1.75, 1.25)\times 10^{-13} \, \mathrm{s}^{-1}$. Using again eq. \ref{eq:UV_photo}, and the specific intensity at $h\nu=12.87 \,\mathrm{eV}$, we get a LW H$_2$ photo-dissociation rate $\Gamma_{\mathrm{UVB,H2}}=2.05\times 10^{-13} \, \mathrm{s}^{-1}$.

By equating the photoionization rates $\Gamma$ and $\Gamma_{\mathrm{UVB}}$, we compute the \quotes{proximity} radius $R_{p}$ within which the flux from the galaxy dominates with respect to the cosmic UVB. We find that, for $\fesc=0.2$, $R_{p}\simeq (250, 168)\, \mathrm{kpc}$ for (H, He), respectively. This implies that the ionization state of the observed outflow, extending to about $10\, \mathrm{kpc}$, is completely governed by the galactic flux. Obviously, if $\fesc=0$ the UVB is the only source of photons. 

\subsection{Cooling function}

Having derived the values of the photoionization and photodissociation rates at each radius, we derive the value of the net (i.e. cooling $-$ heating) cooling function $\Lambda(T,n,r)$ using the data tabulated in \citet[][]{gnedin2012cooling}. Their model includes the effects of different cooling mechanisms, such as metal line cooling, atomic cooling, photoelectric effect on H, He. We also assume that the gas has solar metallicity when evaluating the cooling function. This choice is motivated by simulations of $z=6$ galaxies \citep{pallottini2017, pallottini:2019} and by extrapolating of the mass-metallicity relation for galaxies at $z=6$ \citep{mannucci:2012}.

The results are shown in Fig. \ref{fig:global_cooling} as a function of $T$ for different gas densities, $n$, and two values of the escape fraction, $\fesc=0, 0.2$. As already mentioned, if $\fesc=0$ the ionizing photons are those from the UVB whose intensity at $z=6$ is given by the \citet{haardt2012radiative} model. For $\fesc=0.2$ the cooling function depends explicitly on the radius $r$: for displaying purposes, we fix $r=1\,\mathrm{kpc}$.

There are striking differences between the two $\fesc$ cases. For $\fesc=0.2$ (left panel) we see that the main effect of the strong galactic flux at a distance of 1 kpc is to dramatically depress the ability of the gas to cool in the temperature range $10^{4-6}$ K, particularly for low gas densities. The decrease of the peak is mostly produced by the fact that H (and partly also He) atoms, providing the main cooling channel via the excitation of the Ly$\alpha$ transition, become ionized and therefore unable to radiate efficiently. The equilibrium temperature, given by the condition $\Lambda=0$, is identified by the spikes in the curves, where a transition from a cooling to a heating-dominated regimes at lower $T$ takes place. The equilibrium values range in $\log T=4.1 - 4.7$, with the warmer solutions applying to lower densities.

The situation is considerably different if ionizing radiation from the galaxy is not allowed to escape in the halo ($\fesc=0$, right panel). In this case the much lower intensity of the UVB alone produces only a very limited suppression of the cooling function, and only for low densities, $n < 0.01\, \cc$. Equilibrium temperatures are consistently lower for $\fesc=0$, due to the decreased photoheating provided by the UVB. 

We conclude that the cooling function is heavily dependent on $\fesc$. Given that in turn the observable properties of the outflow, as e.g. its \CII emission, depend strongly on gas temperature, this raises the interesting possibility that outflows might be used to indirectly probe $\fesc$. We will return to this point later on. 

\section{Outflow structure}\label{sec_structure_of_the_outflow}

\begin{figure*}
    \centering
    \includegraphics[width=1.025\textwidth,height=21cm]{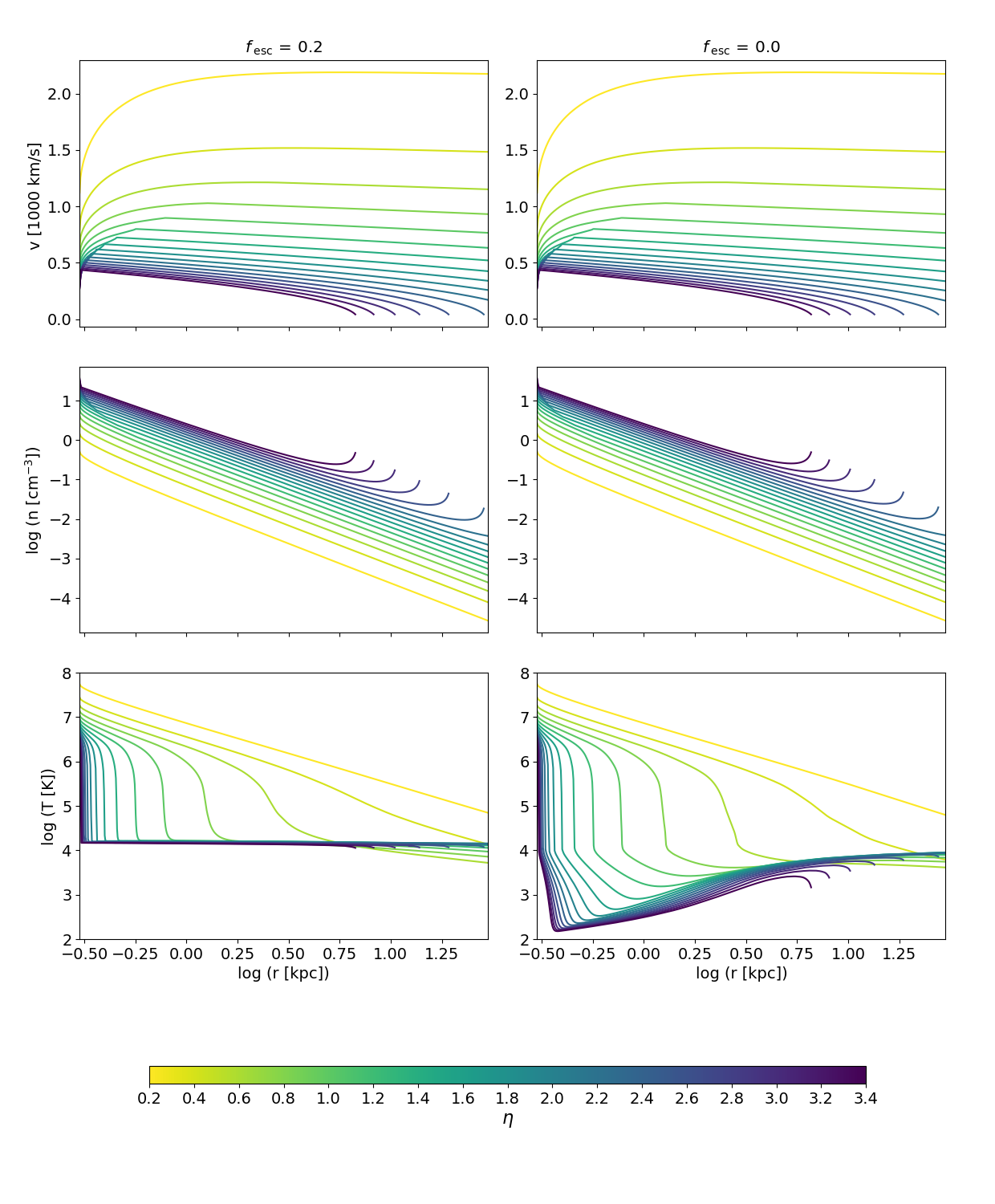}
    \caption{Radial profiles of key outflow thermodynamical variables obtained for the cooling/gravity model (eq. \ref{eq:cooling_model_equations_v}). Shown are the two cases $\fesc=0.2$ (left column), and $\fesc=0$ (right).
    \textit{Top row}: Velocity ($v$). For high values of the mass loading factor $\eta$, gravity slows down the outflow until a stalling radius at which $v=0$ is reached.
    \textit{Middle}: Density ($n$). The radial dependence of the density is generally $n\propto r^{-2}$, but its value increases as the gas slows down due to gravity.
    \textit{Bottom}: Temperature ($T$). Note the different temperature profiles beyond cooling radius. For $\fesc=0$ the outflow cools to lower temperatures and reaches the equilibrium value only at a much larger radii. 
    \label{fig:global_profiles}
    }
\end{figure*}

\begin{figure*}
    \centering
    \includegraphics[width=1.02\textwidth,height=14cm]{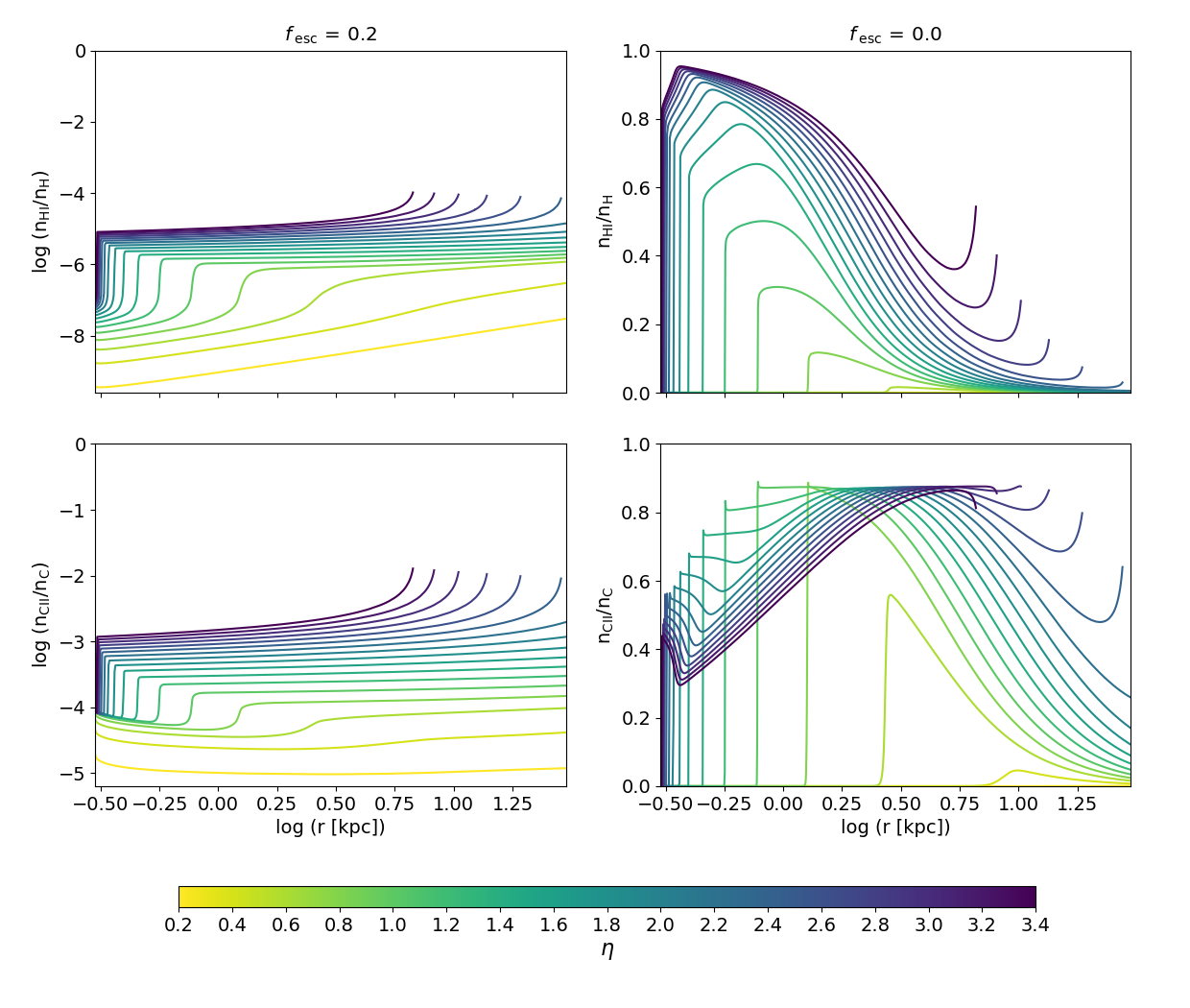}
    \caption{Outflow radial ionization profiles. Shown are the two cases $\fesc=0.2$ (left column), and $\fesc=0$ (right). Note the linear scale in the right panels.
    \textit{Top row}: Neutral hydrogen fraction (eq. \ref{eq:ne}). \textit{Bottom}: Singly ionized carbon fraction from (eq. \ref{eq:densityCII}).
    \label{fig:global_ionization}
    }
\end{figure*}

We present in Fig. \ref{fig:global_profiles} the thermodynamic structure of the outflow as derived from the numerical solution of the hydrodynamical equations (eq.s \ref{eq:cooling_model_equations}). In the following we first discuss the case $\fesc=0.2$, and then consider the case $\fesc=0$.

\subsection{Case for $\fesc=0.2$}

The first column of Fig. \ref{fig:global_profiles} shows the radial profiles of the key hydrodynamical variables, $v, n, T$ for $\fesc=0.2$ for different values of the mass load parameter, $\eta$. 

For $\eta \lsim 1$ the radial asymptotic dependencies are still $v \approx \mathrm{const.}$ and $n\propto r^{-2}$ as in the no gravity, no cooling case. However, when $\eta \gsim 1$ the initial density is high enough for gravity to become important. This reduces the velocity up to a stalling radius, $r_{\rm stop}$, where the velocity drops to zero. The position of the stalling point moves closer to the galaxy as $\eta$ increases. 

Cooling introduces new, striking features in the temperature profiles shown in the lower panels of Fig.  \ref{fig:global_profiles}. For high values of the mass loading factor ($\eta \gtrsim 1$) the gas starts cooling at a distance $r_{\mathrm{cool}}$ that gets smaller as $\eta$ increases. The cooling is quite rapid, and it stops at the equilibrium temperature (see Fig. \ref{fig:global_cooling}) around $10^4 \,\mathrm{K}$. Beyond the cooling radius the outflow is subject to a quasi-isothermal expansion.

\subsection{Case for $\fesc=0$}

We show the radial profiles of the thermodynamic quantities for $\fesc=0$ in the right column of Fig. \ref{fig:global_profiles}, allowing a direct comparison with the $\fesc=0.2$ case.\footnote{For $\fesc=0$ the profiles closely resemble the ones in \citet{Thompson16}, where cooling and gravity are similarly implemented in the \citeCC85 model. The small differences reside in the adopted cooling function model  -- \citet{gnedin2012cooling} vs. \citet{oppenheimer&schaye} -- and on the different UVB photoionization rate at $z=6$ (this work) vs. $z=0$ \citet{Thompson16}.}

The velocity and density profiles are very similar to the ones for $\fesc=0.2$, i.e. they are not significantly affected by the presence of a ionizing galactic flux. On the other hand, the temperature shows a different behaviour beyond $r_{\rm cool}$ as expected from the different shapes of the cooling functions (Fig. \ref{fig:global_cooling}). For $\fesc=0$ the gas is able to cool down to a temperature of a few hundred degrees. At larger radii the gas slowly heats up as the net cooling function takes negative values (i.e. the photoionization heating takes over as density decreases). As we will show in the following paragraph, temperatures of a few $\times 100$K allow a significant presence of \CIIion, and thus a potentially observable \CII emission. 
%

\subsection{Ionization structure}\label{sec_ionization_structure}

From the above density and temperature profiles of the outflow we can now compute the ionization state of different species as a function of the radial distance from the galaxy. We present the details of the ionization equilibrium calculations for H and C in App. \ref{sec_CII_density}.

The resulting ionization radial profiles are shown in Fig.     \ref{fig:global_ionization}. For $\fesc=0.2$ both H and C atoms are largely in the form of \HII and \CIIIion. In particular, the fraction of singly ionized carbon is $x_{\rm CII}=n_{\rm CII}/n_C < 10^{-3}$. In these conditions, \CII line emission is strongly suppressed. For this reason, in the following we will concentrate on the case $\fesc=0$, which gives the most promising results. 

Looking at the right column of Fig. \ref{fig:global_ionization} ($\fesc=0$), we see that these models can produce considerable amounts of \CIIion. As the gas cools to a few hundred degrees K beyond $r_{\rm cool}$ carbon recombines, and $x_{\rm CII} \gsim 0.5$ in most cases but the lowest values of $\eta$. The outflow is essentially neutral as H has also largely recombined. As the outflow temperature increases again towards larger radii \CIIion ions are collisionally ionized, and their abundance decreases, albeit remaining significant. Thus, cooling outflows can potentially explain the observed extension of \CII halos around early galaxies.

\begin{figure*}
    \centering
    \includegraphics[width=1.025\textwidth,height=8.5cm]{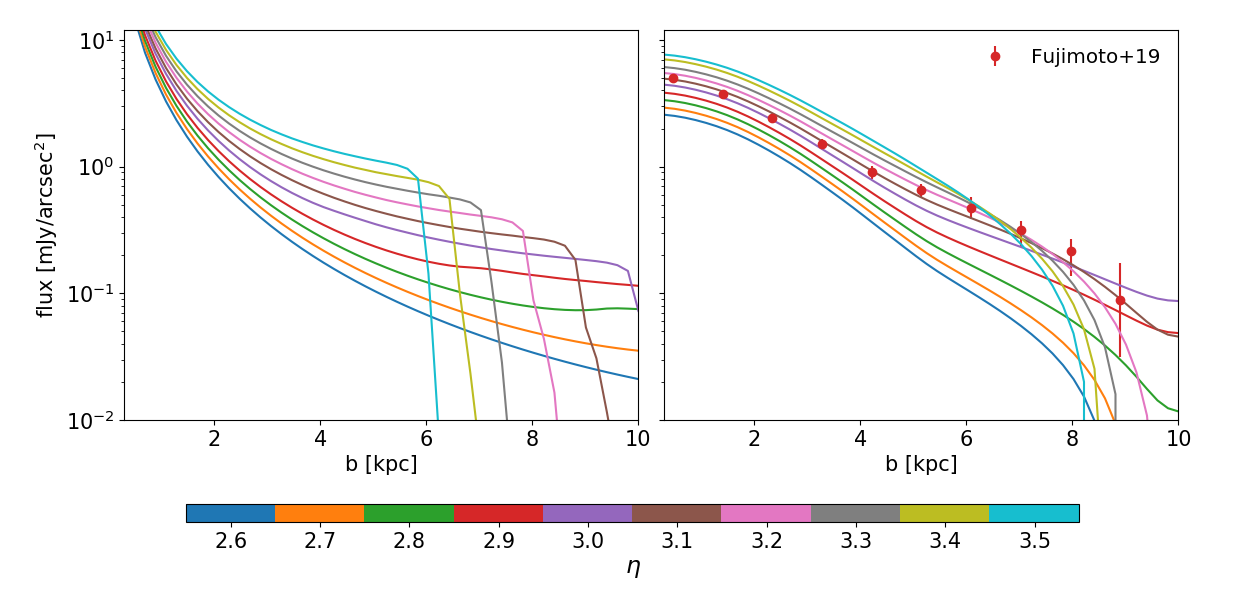}
    \caption{\textit{Left}: Stacked \CII surface brightness profiles for different values of $\eta$ as a function of the impact parameter $b$. Each profile combines the different SFR values of the 18 galaxies considered by \citeF19. The profiles with $\eta\gtrsim3.0$ are discontinuous because the highest values of the SFRs have a stopping radius $r_\mathrm{stop}< 10\, \mathrm{kpc}$. \textit{Right}: comparison of the profiles with data from \citeF19. The profiles are convolved with the same beam as in the observation (shown in Fig. \ref{fig:vesc_emission}).
    \label{fig:global_emission}
    }
\end{figure*}

\section{\CII line emission}\label{sec_CII_emission}

To enable a direct comparison between our model and the observed \CII surface brightness the last step is to derive the expected \CII line emission from the computed $x_\mathrm{CII}$ and $T$ radial profiles.

Similarly to other works \citep{vallini2015,kohandel:2019,ferrara:2019}, we use an analytical model to compute the \CII~line emisssion. We follow \citet{tielens2005book} and write the local \CII emissivity in the low-density regime as
\begin{align}
\Lambda_{\mathrm{CII}} = 2.1 \times 10^{-23}\, A_\mathrm{C} \,\left(1+ 420\,x_\mathrm{CII} \right)\, e^{-92/T} \,\, \mathrm{erg} \,\mathrm{cm}^{3}\, \mathrm{s}^{-1}\,.
\end{align}
The abundance of carbon is taken to be ${A_C}=2.7\times 10^{-4}$  \citep{asplund2009}. Since \CII~emission is typically optically thin \citep{osterbrock1992}, the \CII surface density, $\Sigma_\mathrm{CII}$, along a radial line of sight is simply obtained by integrating the emissivity,
\begin{align}
\Sigma_\mathrm{CII}(r)=\int n^2(r) \,\Lambda_{\mathrm{CII}}(T(r))\,\d r
\label{Int}
\end{align}
It is useful to express $\Sigma_\mathrm{CII}$ as function of the impact parameter $b$, i.e. the distance between the line of sight and the centre of the galaxy.  Eq. \ref{Int} can then be written as
\begin{align}
\Sigma_\mathrm{CII}(b)&=\int_{-\infty}^{+\infty} n^2(r(x))\, \Lambda_{\mathrm{CII}}(r(x))\, \d x = \nonumber\\
&= 2 \int_{b}^{+\infty}  n^2(r)\, \Lambda_{\mathrm{CII}}(T, n, r)\, \frac{r}{\sqrt{r^2 - b^2}}\,\d r \,.
\label{eq:final_intensity}
\end{align}

\section{Comparison with data}\label{sec_comparison_with_data}

\begin{figure*}
    \centering
    \includegraphics[width=1.025\textwidth,height=8.5cm]{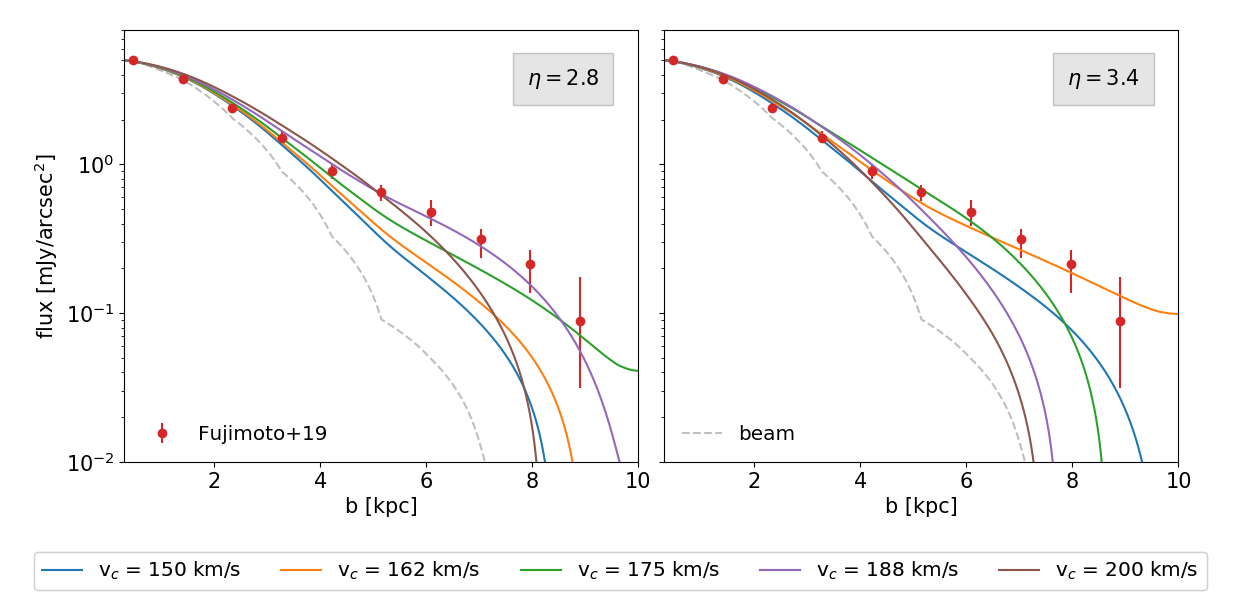}
    \caption{Predicted \CII surface brightness profiles (solid lines) as a function of the impact parameter $b$ for different values of the centrifugal velocity $v_\mathrm{c}$, compared with the data from \citeF19 (points). Two values of the mass loading factor are shown: $\eta=2.8$ (left panel), $\eta=3.4$ (right). The profiles are normalised to the central value of the data, and convolved with the same ALMA beam (grey dashed line) used by \citeF19.
    \label{fig:vesc_emission}
    }
\end{figure*}

We want to use our results to interpret \citeF19 results, which are obtained from a stacking of the sample including galaxies with different star formation rates (with a mean SFR $= 40 \pm 5\,\msun {\rm yr}^{-1}$). The SFR linearly affects our boundary conditions (eq. \ref{eq:bc}), and thus it has a relevant effect on the variables profiles and on our final prediction for the \CII emission. Therefore, in order to perform a fair comparison with observations, we take the SFR value of every single galaxy considered in \citeF19, and use it to compute the \CII emission. The individual galaxy predictions are then stacked into a single profile, which is still a function of $\eta$.
More rigorously, we compute
\begin{align}
\Sigma_\mathrm{CII}(b;\eta)= \frac{1}{N}\sum_i \Sigma_\mathrm{CII}(b;\mathrm{SFR}_i, \eta)\,, \label{eq:stacked_intensity}
\end{align}
where $N=18$ is the number of galaxies considered in the \citeF19 sample, and $\Sigma_\mathrm{CII}(b;\mathrm{SFR}_i, \eta)$ [erg cm$^{-2} {\rm s}^{-1}$] is given in eq. \ref{eq:final_intensity}. 

For a direct comparison with the results in \citeF19, we convert the \CII surface density in a surface brightness (i.e. flux per unit solid angle), measured in $\mathrm{mJy\,arcsec}^{-2}$. We do this dividing $\Sigma_\mathrm{CII}$ by the observed \CII linewidth $\Delta\nu_\mathrm{obs}$:
\begin{align}
\Delta\nu_\mathrm{obs} = \frac{\Delta {\rm v}}{c} \frac{\nu_0}{1+z}\,,
\end{align}
where $\nu_0=1900\,\mathrm{GHz}$ is the restframe frequency of the \CII line. From \citeF19:
\begin{align}
\Delta {\rm v} \equiv \mathrm{FWHM} = 296\pm40\kms
\end{align}
Since the luminosity per unit frequency and per unit solid angle of the \CII line can be written as: 
\begin{align}
\frac{\d\mathrm{L}_\mathrm{CII}}{\d \Omega \, \Delta \nu_\mathrm{obs}} = \frac{\Sigma_\mathrm{CII}}{\Delta \nu_\mathrm{obs}}d_A^2\,,
\end{align}
the flux per unit solid angle is then:
\begin{align}
\frac{\d{\cal F}}{\d\Omega} = \frac{\Sigma_\mathrm{CII}}{\Delta \nu_\mathrm{obs}}\frac{d_A^2}{4\pi d_L^2 } = \frac{\Sigma_\mathrm{CII}}{4\pi\Delta \nu_\mathrm{obs}(1+z)^4};
\end{align}
for the \citeF19 sample we use the average redshift $\langle z \rangle=6$. 

\begin{figure}
    \centering
    \includegraphics[width=0.5\textwidth]{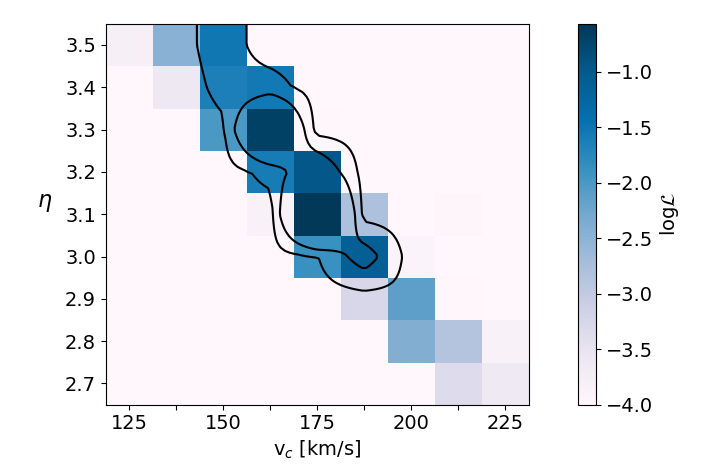}
	\caption{Likelihood $\mathcal{L}(x; \eta, v_c)= \exp[-\chi^2(x;\eta,v_c)/\mathrm{ndof}]$ of the model to the \citeF19 data as a function of the two free parameters, $\eta$ and $v_c$. The black contours represent the 68\% (inner line) and 95\% (outer) confidence levels.
	\label{fig:chi_square}
	}
\end{figure}

We plot the most interesting ($\eta\geq 2.6$) flux profiles $[\mathrm{mJy\,arcsec}^{-2}]$ in the left panel of Fig. \ref{fig:global_emission}. As it is clear from the plot, stacking the flux results in profiles with significant discontinuities. This is because an increase in SFR produces a brighter emission, but at the same time the wind is slowed down at smaller stalling radii.\footnote{The dependence of the stalling radius, $r_\mathrm{stop}$, on the SFR can be inferred from eq. \ref{eq:cooling_model_equations_v}. For temperatures $T\approx10^{2-4}\, \mathrm{K}$, the flow is highly supersonic. Also, the cooling length of the gas remains larger than the outflow extent. Hence, using these approximations it is straightforward to obtain an analytical solution for the profile $v(r)$ (see also \citet{Thompson16}), from which it follows that
\begin{align}
r_\mathrm{stop} \approx r_\mathrm{cool}\,e^{v^2(r_\mathrm{cool})/2v_c^2},
\end{align}
which implies that $r_\mathrm{stop}$ increases with the cooling radius, $r_\mathrm{cool}$. As the latter has a strong inverse dependence on the outflow rate $\eta {\rm SFR}$ \citep{Thompson16}, a higher SFR value results in a smaller $r_\mathrm{stop}$.} Hence, beyond $r_\mathrm{stop}$ the emission drops to zero. 

For a proper comparison with observations, we convolve our profiles with the ALMA beam used in the observation runs (shown in Fig. \ref{fig:vesc_emission} with a grey dashed line). This procedure smooths out the expected discontinuities. The final prediction for the observed \CII line surface densities, $\Sigma_{\rm CII}$, as a function of impact parameter is shown in the right panel of Fig. \ref{fig:global_emission}. By looking at Fig. \ref{fig:global_emission}, we conclude that the profiles with $\eta\gsim 2.6$ result in a surface brightness broadly consistent with those observed by \citeF19, with central values in the range $\approx 1-5\,\mathrm{mJy\,arcsec}^{-2}$.
The profiles with the highest loading factors, $3.2 < \eta < 3.4$, are characterised by a very high \CII surface brightness in the central regions of the halo, but they drop abruptly at the stalling radius, $r_\mathrm{stop}$, which is smaller than the observed extension of the \CII emitting halo. Less mass-loaded outflows ($\eta=2.6-3.0$) have a low $\Sigma_{\rm CII}$, but they extend out to $r > 10\,\mathrm{kpc}$.

The solution that best fits the data represents a compromise between these two trends. By performing a $\chi^2$ fitting procedure, we find that the best solution ($\chi^2/\mathrm{ndof}= 8/10$) is the one with $\eta=3.1$. We conclude that our model predicts the observed emission with a satisfying level of accuracy.

A mass loading factor $\eta=3.1$ corresponds to an outflow rate $\dot{M}_\mathrm{out}=4\pi v \rho r^2\approx 125 \,\msun {\rm yr}^{-1}$. The implied total mass of gas (carbon) in the halo is $6.5\times 10^9\,\msun$ ($1.7\times 10^6\,\msun$). The outflow rate resulting from our analysis is higher than (but still consistent at 3$\sigma$ with) the one found in \citet{gallerani:2018}, i.e. $\dot{M}_\mathrm{out} = 56\pm23 \,\msun {\rm yr}^{-1}$. These authors detected the presence of \CII line broad ($\approx 500 \kms$) wings indicative of outflows by stacking nine $z\approx 5.5$ galaxies, part of the \cite{capak2015} sample, with a mean SFR $= 31 \pm 20\,\msun {\rm yr}^{-1}$, namely slightly lower than the F19 sample.

\subsection{Dependence on halo circular velocity}\label{sec_v_esc0}

As a final step, we explore the dependence of the results on the dark matter halo circular velocity, $v_c$ (eq. \ref{eq:v_esc}). We select for the analysis two values ($\eta=2.8,\,3.4$) close to the best-fitting value $\eta=3.1$ found above, and look at the \CII surface brightness profiles for different values of $v_c$. We normalize the profiles to the central value of the \citeF19 data to emphasize the differences in the profile shapes.

The results are shown in Fig. \ref{fig:vesc_emission}. In each panel, only one curve satisfactorily matches the data. For $\eta =2.8$ ($\eta=3.4$) an excellent fit is obtained for $v_c=188\kms$ ($v_c=162\kms$). These values correspond to dark matter halo masses around $10^{11}\,\msun$. 

It is useful to comment on the dependence of \CII emission on $\eta$ and $v_c$. While $\eta$ affects primarily the overall halo brightness by regulating the outflow density, changing $v_c$ is equivalent to modify the strength of the gravitational field. As it is clear from Fig. \ref{fig:vesc_emission}, a deep gravitational potential ($v_c \gtrsim 200\kms$) results in values of $r_{\rm stop}$ which do not match the observed extension of the emitting halo. Weaker potentials ($v_c \simlt 150\kms$) are instead unable to slow down the outflow and therefore maintain a sufficiently high gas density in the outer regions of the halo. In this case, the low density of the gas results in a very faint (undetectable) emission. In addition, the low density gas is more susceptible to photoionization by the galactic and/or cosmic UV field turning \CIIion into \CIIIion. Such key role of the gravitational confinement has been noted also in recent hydrodynamical simulations results \citep{li2019supernovae}.

In order to further generalize our results, in Fig. \ref{fig:chi_square} we have performed a full parameter study for $\eta$ and $v_c$. We take $\eta$ ranging from $2.7$ to $3.5$ and $v_c$ ranging from $125$ to $225\kms$. For every couple of parameters, we compute the predicted $\Sigma_{\rm CII}$ profile, and compute the likelihood of the model to the \citeF19 data as in the previous cases. The resulting likelihood function is shown in Fig. \ref{fig:chi_square}. 

Generally, a tight anti-correlation between $\eta$ and $v_c$ is found, but the likelihood shows a narrow maximum around the values close to the ones identified previously, i.e. $\eta = 3.2 \pm 0.10$ (or $\dot{M}_\mathrm{out}= 128 \,\msun {\rm yr}^{-1}$) and $v_c = 170 \pm 10 \kms$. These results imply that extended halos might be used to set constrains on the mass loading factor and dark matter halo mass of early galaxies.

\section{Summary and Conclusions} \label{sec_conclusions}

We have proposed that the recently discovered \citep{Fujimoto19}, very extended ($\approx 10$ kpc) \CII emitting halos around EoR galaxies are the result of supernova-driven cooling outflows. Our model contains two parameters, the outflow mass loading factor, $\eta=\dot{M}_\mathrm{out}$/SFR, and the parent galaxy dark matter halo circular velocity, $v_c$. The outflow model successfully matches the observed \CII surface brightness if $\eta = 3.20 \pm 0.10$ and $v_c = 170 \pm 10\kms$. Given that for the F19 sample the mean SFR $= 40 \pm 5\,\msun {\rm yr}^{-1}$, the predicted outflow rate is $\dot{M}_\mathrm{out}=128\pm 5\,\msun {\rm yr}^{-1}$.  We also note that the presence of extended \CII halos requires a ionizing escape fraction from the parent galaxy $\fesc \ll 1$. Values of $\fesc \gsim 0.2$, as those required by most reionization models, produce halo UV fields that are too intense for \CII to survive photoionization. 

The success of the model largely relies on the fact that we follow precisely the catastrophic cooling of the outflow occurring within the central kpc. We find that cooling takes place for conditions (gas density $n\approx 1 \,\mathrm{cm}^{-3}$, temperature $T\approx 10^6\,\mathrm{K}$) consistent with the ones found by previous models and simulations \citep{Thompson16, Scannapieco:2017, gray2019catastrophic}. The gas cools very rapidly to $T \approx {\rm few} \times 100$ K, at the same time recombining. In this regime the formation and survival of \CIIion ions is guaranteed. CII ions are transported by the neutral outflow at velocities of 300-500 $\kms$. In brief, \CII halos, according to our model, are the result of cold neutral outflows from galaxies.

Although the model has been applied here to stacked data, it can be readily adapted to individual high-$z$ galaxies, as those observed e.g. by the ALMA ALPINE survey, which are now becoming available (\citealt{ginolfi:2019}; see also \red{\citet{fujimoto:2020}}). The model returns key information on early galaxies, such as (i) the presence of outflows and their mass loading factor/outflow rate; (ii) the dark matter halo mass; (iii) the escape fraction of ionizing photons. These are are all crucial quantities which are hardly recovered from alternative methods at high redshifts. By modelling galaxies on an individual basis it will be also possible to clarify whether the emission profile and extension of the \CII halo is related to the SFR of the galaxy. 

Clearly, the fact that the extended \CII halos surface brightness can be successfully fit by our model does not guarantee that outflows are the only possible explanation. Alternative interpretations, such as the presence of satellites, also need to be carefully explored. Interestingly, \citet{gallerani:2018} reported evidence for starburst-driven outflows in nine $z\approx 5.5$ galaxies from the presence of broad wings in the \CII line. Although they could not exclude that part of this signal is due to emission from faint satellite galaxies, their analysis favoured the outflow hypothesis.

Remarkably, although two independent hydrodynamical zoom-in simulations \citep{pallottini2017b, Arata:2019} have successfully matched both the dust and stellar continuum profiles deduced from \citeF19 observations, the same simulations could not reproduce the extended \CII line emission. This might be due to an incomplete treatment of stellar feedback, or to numerical resolution issues related to the outflow catastrophic cooling. Our simple model is instead able to perfectly match the observed surface brightness. Hence, insight can be likely gained from a detailed comparison with simulations.

Alternatively, the failure of the simulations might indicate that the additional energy input required to transport the gas at such large distances could be provided by an AGN. Although the inferred value of $\eta=3.2$ is marginally consistent with starburst-driven outflows \citep[e.g.][]{Heckman15}, it is probably more typical of AGN \citep{Fiore_2017}. This hypothesis must be tested via dedicated hydrodynamical simulations including radiative transfer. 

In spite of its success, the model presented here contains several limitations and hypothesis that will need to be removed in the future. The present one-dimensional treatment should be augmented with a full 3D numerical simulation of the outflow, also dropping the steady state assumption made here. A more realistic treatment of the circumgalactic environment is also necessary. Simulations show that accounting for an external CGM pressure might result in the formation of shocks in the outflowing gas \citep{samui:2008, Lochaas:2020, gray2019catastrophic}. Although we do not expect these shocks to dramatically affect the derived overall outflow structure, the detailed profile and extension of the [CII] emitting region might turn out quantitatively different. This can be tested with less idealized, 3D simulations. Non-equilibrium cooling/recombination effects should be considered when computing ionic abundances. Finally, the effects of CMB on \CII emission \citep{dacunha2013,pallottini2017b,kohandel:2019}, particularly in the external, low-density regions of the outflow must be included in the calculation. Although some of these improvements might affect the quantitative conclusions of this paper, it appears that so far outflows remain the best option to explain the puzzling nature of extended \CII halos. These systems might be the smoking gun of the process by which the intergalactic medium was enriched with heavy elements during the EoR, as witnessed by quasar absorption line experiments \citep{Dodorico13,Meyer19,Becker19}.

\section*{Acknowledgements}
EP thanks Edoardo Centamori for useful suggestions.
AF acknowledges support from the ERC Advanced Grant INTERSTELLAR H2020/740120.
LV acknowledges funding from the European Union's Horizon 2020 research and innovation program under the Marie Sk\l{}odowska-Curie Grant agreement No. 746119. 
SF acknowledges the Cosmic Dawn Center of Excellence funded by the Danish National Research Foundation under then grant No. 140.
This research was supported by the Munich Institute for Astro- and Particle Physics (MIAPP) of the DFG cluster of excellence \quotes{Origin and Structure of the Universe}.
Partial support from the Carl Friedrich von Siemens-Forschungspreis der Alexander von Humboldt-Stiftung Research Award is kindly acknowledged (AF).
We acknowledge use of the Python programming language \citep{VanRossum1991}, Astropy \citep{astropy}, Matplotlib \citep{Hunter2007}, NumPy \citep{VanDerWalt2011}, and SciPy \citep{scipyref}.


\bibliographystyle{mnras}
\bibliography{biblio,codes} 


\appendix

\section{CII density}\label{sec_CII_density}

In order to predict the \CII line emission from the outflow it is necessary to evaluate the fraction of carbon found in the singly ionized state. We start by assuming that the electron density is equal to the proton density, $n_e \approx n_p$, i.e. we neglect contributions from other ionized species, such as He and C because of their lower abundance and/or higher (for He) ionization potential. Then we write the hydrogen ionization equation 
\begin{align}
n_\mathrm{H} \Gamma_{\mathrm{H}} + n_\mathrm{H} n_e k_{\mathrm{H}} = n_e\, n_{p} \,\eta_{\mathrm{H}}
\label{ionbal}
\end{align}
where $\Gamma_\mathrm{H}$, $k_\mathrm{H}$, and $\alpha_\mathrm{H}$ are the hydrogen photoionization, collisional ionization, and recombination coefficients respectively. For $\Gamma_\mathrm{H}$ we use the expresions given in Sec. \ref{Radfields}; $k_\mathrm{H}$ is taken from \citet[][Appendix B]{bovino:2016aa}; for $\alpha_H$ we use the power-law approximation to Case B radiative recombination given by \citet{tielens2005book},
\begin{align}
\eta_\mathrm{H} = 4.18 \times 10^{-13} \, \bigg(\frac{T}{10^4 \, \mathrm{K}}\bigg)^{-0.75} \,\, \mathrm{cm}^3\, \mathrm{s}^{-1}.
\label{eq:beta_H}
\end{align}
Using $n_\mathrm{p} + n_\mathrm{H} = A_\mathrm{H} n$, where $A_\mathrm{H}$ is the cosmic hydrogen abundance\footnote{We assume a solar chemical composition \citep{asplund2009} for which $A_{\mathrm{H}} = 0.76, A_{\mathrm{C}} = 2.69 \times 10^{-4}$.}, $n$ the total gas density, and defining $x_\mathrm{e}=n_\mathrm{e}/n$, we can recast eq. \ref{ionbal} in the following form:
\begin{align}
(\eta_\mathrm{H}+k_\mathrm{H})\,x_e^2 + \bigg(\frac{\Gamma_{\mathrm{H}}}{A_H n}-k_\mathrm{H}\bigg)\,x_e - \frac{\Gamma_\mathrm{H}}{A_H n}= 0,
\label{eq:ne}
\end{align}
from which the H ionization fraction can be obtained. 

We now turn to carbon and write the equivalent ionization equations assuming a detailed balance among three states, with number density $n_{\mathrm{CI}}$, $n_{\mathrm{CII}}$, $n_{\mathrm{CIII}}$, of C atoms ionization, 
\begin{subequations}
\begin{align}
n_\mathrm{CI} \Gamma_{\mathrm{CI}} + n_\mathrm{CI} n_e k_{\mathrm{CI}} &=n_e\, n_\mathrm{CII} \,\eta_{\mathrm{CII}};\\
n_\mathrm{CII} \Gamma_{\mathrm{CII}} + n_\mathrm{CII} n_e k_{\mathrm{CII}}&=n_e\, n_\mathrm{CIII} \,\eta_{\mathrm{CIII}}.
\end{align}
\end{subequations}
The photoionization, collisional ionization, and recombination coefficients are $\Gamma_{\mathrm{CI}}$, $\Gamma_{\mathrm{CII}}$, $k_\mathrm{CI}$, $k_\mathrm{CII}$, and $\alpha_{\mathrm{CII}}$, $\alpha_{\mathrm{CIII}}$, respectively. With the bound $n_\mathrm{CI}+n_\mathrm{CII}+n_\mathrm{CIII}=A_\mathrm{C} n \equiv n_\mathrm{C}$, we can solve the equations above and obtain ionization fraction of Carbon $x_\mathrm{CII}= n_\mathrm{CII}/n_\mathrm{C})$:
\begin{align}
x_\mathrm{CII} = \bigg(1+\frac{\Gamma_{\mathrm{CII}}}{n_e\,\eta_{\mathrm{CIII}}}+\frac{n_e\,\eta_{\mathrm{CII}}}{\Gamma_{\mathrm{CI}}+n_e k_\mathrm{CI}}+ \frac{k_\mathrm{CII}}{\eta_\mathrm{CIII}}\bigg)^{-1} \label{eq:densityCII}
\end{align}

The photoionization rates $\Gamma_{\mathrm{CI}}$ and $\Gamma_{\mathrm{CII}}$ can be computed in the same way as done for H and He (eq. \ref{eq:photo_profiles}) using the photoionization cross section data in Table \ref{tab:params}. We finally get
\begin{subequations}
\begin{align}
\Gamma_{\mathrm{CI}}(r)&=7.5\times 10^{-7}\, \bigg(\frac{\mathrm{kpc}}{r}\bigg)^2 \fesc \,\, \mathrm{s}^{-1}\\
\Gamma_{\mathrm{CII}}(r)&=1.85\times 10^{-8}\, \bigg(\frac{\mathrm{kpc}}{r}\bigg)^2 \fesc \,\, \mathrm{s}^{-1}\,,
\end{align}
\end{subequations}
and the analogous quantities for the case $\fesc=0$ in which the only radiation field is the UVB taken from \citet{haardt2012radiative} and the parameters in Table \ref{tab:params}. We obtain:
\begin{subequations}
\begin{align}
\Gamma_{\mathrm{UVB,CI}}&=1.34\times 10^{-12} \,\, \mathrm{s}^{-1}\\
\Gamma_{\mathrm{UVB,CII}}&=6.77\times 10^{-14} \,\, \mathrm{s}^{-1}
\end{align}
\end{subequations}

Recombination rates must include both radiative and dielectronic recombination. For these we use the following approximations \citep{tielens2005book}: 
\begin{subequations}
\begin{align}
\alpha_\mathrm{CII} &= 10^{-13} \bigg[ 4.66 \bigg(\frac{T}{10^4 \, \mathrm{K}}\bigg)^{-0.62} + 1.84  \bigg]\,\, \mathrm{cm}^3\, \mathrm{s}^{-1},\\
\alpha_\mathrm{CIII} &=10^{-12} \bigg[ 2.45  \, \bigg(\frac{T}{10^4 \, \mathrm{K}}\bigg)^{-0.65} + 6.06 \bigg]\,\, \mathrm{cm}^3\, \mathrm{s}^{-1}.
\end{align}
\end{subequations}
Finally, the collisional ionization rates, $k_\mathrm{CI}$ and $k_\mathrm{CII}$, are taken from \citet[][Table 1]{voronov_practical_1997}.

\bsp	
\label{lastpage}

\end{document}

%% file: additional_tex/pack.tex
\usepackage{newtxtext,newtxmath}

\usepackage[T1]{fontenc}
\usepackage{ae,aecompl}

\usepackage{graphicx}	
\usepackage{amsmath}	
\usepackage{amssymb}	
\usepackage{comment}
\usepackage{siunitx}

\usepackage{xcolor}

%% file: additional_tex/definitions.tex
\def\be{\begin{equation}}
\def\ee{\end{equation}}
\newcommand\code[1]{\textsc{\MakeLowercase{#1}}}

\newcommand\quotes[1]{``{#1}"}
\def\gsim{\lower.5ex\hbox{\gtsima}} 
\def\lsim{\lower.5ex\hbox{\ltsima}} 
\def\gtsima{$\; \buildrel > \over \sim \;$} 
\def\ltsima{$\; \buildrel < \over \sim \;$} \def\gsim{\lower.5ex\hbox{\gtsima}} 
\def\lsim{\lower.5ex\hbox{\ltsima}} 
\def\simgt{\lower.5ex\hbox{\gtsima}} 
\def\simlt{\lower.5ex\hbox{\ltsima}}

\def\msun{{\rm M}_{\odot}}

\def\cc{\rm cm^{-3}}
\def\kms{\,\rm km\,s^{-1}}

\def\fesc{f_{\rm esc}}

\def\S*{$\Sigma_{\rm SFR}$}

\def\CII{\hbox{[C~$\scriptstyle\rm II $]~}}

\def\OIII{\hbox{[O~$\scriptstyle\rm III $]~}}

\def\HH{\hbox{H$_2$}~} 
 
\def\HII{\hbox{H~$\scriptstyle\rm II\ $}} 
\def\CIion{\hbox{C~$\scriptstyle\rm I $~}}
\def\CIIion{\hbox{C~$\scriptstyle\rm II $~}}
\def\CIIIion{\hbox{C~$\scriptstyle\rm III $~}}

\definecolor{apcolor}{HTML}{b3003b}
\definecolor{afcolor}{HTML}{800080}
\definecolor{lvcolor}{HTML}{DF7401}
\definecolor{mdcolor}{HTML}{01abdf} 
\definecolor{cbcolor}{HTML}{ff0000}
\definecolor{sccolor}{HTML}{cc5500} 
\definecolor{sgcolor}{HTML}{00cc7a}

%% file: additional_tex/additional_def.tex
\renewcommand{\d}{\mathrm{d}}
\newcommand{\der}[3][]{\frac{\d ^{#1}#2}{\d {#3}^{#1}}}

\let\oldnabla\nabla
\renewcommand{\nabla}{\vec{\oldnabla}}

\newcommand{\red}[1]{{{\textcolor{red}{\textbf{#1}}}}}